\documentstyle[aps,epsf]{revtex}
\def\`#1{\if#1i{\accent18 \i}\else{\accent18 #1}\fi}
\def\'#1{\if#1i{\accent19 \i}\else{\accent19 #1}\fi}

\begin{document}
\draft

\date{\today}
\title{Influence of the LPM effect and dielectric suppression on
particle air showers }
\author{A.~N.~Cillis\cite{email}, H.~Fanchiotti, C.~A.~Garc\'ia
Canal and S.~J.~Sciutto}
\address{Laboratorio de F\'isica Te\'orica\\
Departamento de F\'isica\\
Universidad Nacional de La Plata\\
C. C. 67 - 1900 La Plata\\
Argentina}

\maketitle

\begin{abstract}
An analysis of the influence of the Landau-Migdal-Pomeranchuk (LPM)
effect on the development of air showers initiated by astroparticles
is presented. The theory of Migdal is studied and compared with other
theoretical methods, particularly the Blankenbecler and Drell
approach.  By means of realistic computer simulations and using
algorithms that emulate Migdal's theory, including also the so-called
dielectric suppression, we study the behavior of the relevant
observables in the case of ultra-high energy primaries. We find that
the LPM effect can significantly modify the development of high energy
electromagnetic showers in certain cases.
\end{abstract}


\section{Introduction}

The study of atmospheric showers initiated by high energy
astroparticles plays a central role in contemporary cosmic ray physics
\cite{Gaisser}. The most important component, in number of particles,
of the air shower is by far the electromagnetic one. Additionally, many
shower observables that give direct information about the properties
of the primary depend strongly on its behavior.

Electron bremsstrahlung and pair production
are the dominant processes in the electromagnetic shower at very high
energies. The standard description of these processes, i.e., the
Bethe-Heitler equation \cite{BH}, can be incomplete at very high
energies due to some effects that drastically reduce the corresponding
cross sections \cite{Klein}. Those mechanisms play a relevant role in
the development of air showers because they can lengthen them, and
consequently move the position of the shower maximum deeper into the
atmosphere.

In the present work we study two of the several suppression processes
that can affect the high energy electromagnetic interactions
\cite{Klein}, namely, the Landau-Pomeranchuk-Migdal (LPM) effect
\cite{LP,Migdal} and the dielectric suppression \cite{Migdal,TM}. The
first of these effects is due to the multiple scattering while the
second one is due to the interaction of the bremsstrahlung photons
with the atomic electrons in the medium, through forward Compton
scattering.

The LPM effect was studied semi-classically by Landau and Pomeranchuk
\cite{LP} and the first quantum mechanical approach was given by
Migdal \cite{Migdal}. Recently, the experimental confirmation of the
LPM effect at the Stanford Linear Accelerator Center (SLAC)
\cite{Klein,SLAC} has originated new theoretical works
\cite{BD,B,B97,BK,Saka,BDMS} and several analysis of its consequences,
particularly in cosmic ray physics \cite{Zas}.

The main purpose of our work is the study of the modifications in
the development of the air showers that take place when the LPM effect
and the dielectric suppression are taken into account. In order to do
that we first study the different theoretical approaches of the
LPM effect. In particular, we compare the Migdal formulation
\cite{Migdal} with other approaches that were developed more recently
\cite{Klein,BD,B,B97}. We also analyze the modifications of the
bremsstrahlung and pair production probabilities in the case of our
interest, that is, when the medium is a ``layer'' of air of infinite
thickness and variable density. Our study is, therefore, complementary
to the recently published review by Klein \cite{Klein} which primarily
treats the LPM effect in the case of solid targets of finite size,
like the ones used in the already mentioned SLAC \cite{SLAC}
experiment.

To complete our study we have developed new LPM/dielectric
suppression procedures, and we have installed them in the AIRES air
shower simulation system \cite{Aires}. The AIRES code has then been
used as a realistic air shower simulator to generate the data used to
make our analysis of the influence of the LPM effect on high energy
showers.

It is worthwhile to mention that AIRES represents a set of programs to
simulate atmospheric air showers and to manage all the associated
output data. Most of the physical algorithms of the AIRES system are
based on the realistic procedures of the well-known MOCCA program
\cite{mocca}. AIRES provides some additional features, for example:
The Earth's curvature is taken into account allowing safe operation of
all zenith angles; the simulation programs can be linked to different
alternative hadronic collision models like QGSJET \cite{QGSJET} or
SIBYLL \cite{SIBYLL1,SIBYLL2}; etc.

The version of AIRES used in this work is capable of processing the
following particles: photons, electrons, positrons, muons, pions,
kaons, eta mesons, nucleons and antinucleons, and nuclei up to
$Z=26$. Among all the physical processes that may undergo the shower
particles, the most important from the probabilistic point of view are
included in the simulating engine. Those processes can be classified
as follows \cite{Aires}: (i) {\em Electrodynamical processes:\/} Pair
production and electron-positron annihilation, bremsstrahlung
(electrons and positrons), knock-on electrons ($\delta$ rays), Compton
and photoelectric effects, LPM effect and dielectric suppression
(discussed in this work). (ii) {\em Unstable particle decays,\/} pions
and muons, for instance. (iii) {\em Hadronic processes:\/} Inelastic
collisions hadron-nucleus and photon-nucleus, generally simulated
using the external packages\footnote{The external hadronic packages
simulate nontrivially the physical processes involved in a
hadron-nucleus inelastic collision. Among other features, both QGSJET
and SIBYLL are capable of generating high energy gammas coming from
the decay of baryonic resonances, even if the probability of such
diffractive processes diminishes as long as the energy of the
projectile particle is increased \cite{sjshadr}.}  (QGSJET or
SIBYLL). Photonuclear reactions, elastic and inelastic. (iv) {\em
Propagation of particles:\/} All particles --even the short lived
ones-- are properly tracked taking always into account the
corresponding mean free paths for the processes they can
undergo.\footnote{This is particularly important for the case of
electromagnetic showers initiated by hadronic primaries. The
electromagnetic component comes mainly from $\pi^0$ decays, and even
if they are short lived particles, the probability of undergoing
hadronic collisions cannot be neglected at ultra-high energies.}
Continuous energy looses (ionization), multiple Coulomb scattering and
geomagnetic deflection are taken into account for charged particles.

 This paper is organized as follows: In section \ref{LPMth} we
introduce the different approaches to the LPM effect and compare the
Migdal formulation with other ones. In section \ref{PI} we explain the
practical implementation of the LPM effect and the dielectric
suppression into the AIRES program.  In section \ref{sectsim} we
analyze the results of the air showers simulations performed with such
code. Finally, in section \ref{conclu} we place our conclusions and
comments.
\section{Theory of the LPM effect}
\label{LPMth}

The so-called LPM effect takes place when an ultra-relativistic
particle emits low energy bremsstrahlung photons passing through
matter. In this case, fewer photons are emitted than predicted for the
case of isolated atoms \cite{BH}. In a similar way, the cross section
for electron-positron pair production is reduced in the case of high
energy gamma rays passing through matter.

 This effect was first predicted by Landau and Pomeranchuk \cite{LP}
some 40 years ago. They treated the classical radiation of a
high-energy particle in a fluctuating random field inside an
infinitely thick medium. Afterwards, Migdal \cite{Migdal} provided the
corresponding quantum mechanical theory giving analytical expressions
for the bremsstrahlung and pair production cross sections when matter
is present. Migdal theory was developed for an infinite target
thickness. More recently, an experiment performed at SLAC \cite{SLAC}
measured the LPM effect finding that there is an acceptable agreement
between the experimental data and Migdal theory which is presently
considered the standard treatment.

The experimental confirmation of the LPM effect triggered new
theoretical works \cite{BD,B,B97,BK,Saka,BDMS}. In particular,
Blankenbecler and Drell \cite{BD}, Blankenbecler \cite{B,B97} and
Klein \cite{Klein} have reanalyzed the problem, upon better
approximations than in the earlier works, and they were able to
consider effects of finite target thickness. We include in this paper
a comparative analysis between different approaches.

To make our paper self-consistent, we start with a qualitative
description of the LPM effect explaining the main features of Migdal
theory. We then compare the final Migdal results with those of
Blankenbecler and Drell.

\subsection{Bremsstrahlung}

Let us  consider first the case of bremsstrahlung where an electron or
positron of energy $E$ and mass $m$ emits a photon of energy $k$ in the
vicinity of a nucleus of charge $Z$.

Neglecting the photon emission angle and the scattering of the electron,
the minimal longitudinal momentum transfer\footnote{The minimal
longitudinal momentum transfered to the nucleus occurs when the final
electron and the photon follow in the same direction of motion than the
initial electron.} to a nucleus, $q_{\parallel}$, in the case when $E \gg
k$ and $E \gg mc^{2}$, is given by \cite{Klein}
\begin{equation}
q_{\parallel}=\frac{km^{2}c^{3}}{2E(E-k)}.   \label{q}
\end{equation}
The formation length can be defined according to the uncertainty
principle: $l_{fo}=\hbar/q_{\parallel}$. $l_{fo}$ is the distance over
which the interaction amplitudes can add coherently.
It is straightforward to show \cite{Klein} that if other interactions are
present while the electron is traversing this distance, then the resulting
amplitude will, in general, be reduced.

Let us consider the case when multiple scattering takes place during the
radiation emission process. Using a small angle approximation, the
longitudinal momentum transfer can be expressed as
\begin{equation}
q_{\parallel}=\frac{km^{2}c^{3}}{2E(E-k)}+\frac{k\theta_{MS/2}^{2}}{2c}
\label{q2}
\end{equation}
where $\theta_{MS/2}$ is the multiple scattering angle in half the
formation length, that is $(E_{s}/E)\sqrt{l_{f}/(2X_{0})}$, with
$E_{s}=mc^{2}\sqrt{4\pi/ \alpha}=21.2$ MeV ($\alpha \simeq 137^{-1}$),
and $X_{0}=\left[4\eta\alpha
r_{e}^{2}Z^{2}\ln(184Z^{-1/3})\right]^{-1}$ is the radiation length
($r_{e}=e^{2}/mc^{2}$ and $\eta$ is the number of atoms per volume
unit).  The effect of multiple scattering becomes significant when the
second term of equation (\ref{q2}) is comparable in magnitude with the
first one.  This is the case when
\begin{equation}
y \lesssim \frac{E}{E+E_{\rm LPM}} \label{rel}
\end{equation}
where
\begin{equation}
y=\frac{k}{E},    \label{y}
\end{equation}
and
\begin{equation}
E_{\rm LPM}= \frac{m^{4}c^{7}X_{0}}{\hbar E_{s}^{2}} \label{Elpm}
\end{equation}  
is a characteristic, material dependent, energy which gives the scale
where the effect cannot be neglected.  For a given energy $E$, the
emission of photons with $y < E/(E+E_{\rm LPM})$ will be affected by
the interference due to multiple scattering; and therefore for $E \gg
E_{\rm LPM}$ the effect extends to the entire photon spectrum $0< y <
1$.

The characteristic energy $E_{\rm LPM}$ can be expressed as
\begin{equation}
E_{\rm LPM}=\frac{m^{2}c^{3}}{16\pi\hbar \eta r_{e}^{2} Z^{2} \ln(184
Z^{-1/3})}.
\label{Elpm2}
\end{equation}
This equation makes evident the fact that $E_{\rm LPM}$ diminishes when the
density of the medium increases. Therefore, for dilute media, the LPM
effect will be appreciable only for energies much higher than the typical
ones used in experiments with dense targets \cite{Klein,SLAC}. In fact,
for air in normal conditions $E_{\rm LPM} \cong 2.2 \times 10^{8}$ GeV; one
can compare this energy with the energy of the electron beam of 8 and 25
GeV used at SLAC \cite{Klein,SLAC}.

The formation length, when multiple scattering is the dominating
 process, can be put in the form
\begin{equation}
 l_{f}=l_{fo}\sqrt{\frac{kE_{\rm LPM}}{E(E-k)}}, \label{lf2}
\end{equation}
To measure the strength of the effect it is convenient to introduce the
suppression factor through
\begin{equation}
S=\frac{d\sigma_{\rm LPM}/dk}{d\sigma_{BH}/dk}  \label{S}
\end{equation}
where $d\sigma_{BH}/dk$ stands for the bremsstrahlung cross
section given by the classical theory of Bethe and Heitler \cite{BH}.
The fact that the cross sections are proportional to the formation
length gives rise to
\begin{equation}
S\cong\sqrt{\frac{kE_{\rm LPM}}{E(E-k)}}  \label{S2}
\end{equation}

For $k$ small in comparison with $E$, The cross section found by Bethe
and Heitler \cite{BH} is proportional to $1/k$. If multiple
scattering is taken into account, this proportionality changes, due to
equation (\ref {S2}), to $1/\sqrt{k}$.

In the Migdal theory of the LPM effect \cite{Migdal}, the multiple
scattering is treated as a diffusion mechanism. The average radiation per
collision and the interference between the radiation from different
collisions are then computed. When collisions occur too close together,
destructive interference reduces the radiation. The multiple scattering  
is treated using the Fokker-Planck technique to solve the
corresponding Boltzmann transport equation. Migdal includes quantum
effects, such as electron spin and photon polarization, but his
calculations only apply for a target of infinite thickness.
The resulting cross section for the bremsstrahlung process reads:
\begin{equation}
\frac{d\sigma _{\rm LPM}}{dk}=\frac{4\alpha r_{e}^{2}\xi (s)}{3k}%
\{y^{2}G(s)+2[1+(1-y)^{2}]\Phi (s)\}Z^{2}\ln\left(
\frac{184}{Z^{\frac{1}{3}}}\right)
    \label{Blpm}
\end{equation}
where
\begin{equation}
s=\sqrt{\frac{kE_{\rm LPM}}{8E(E-k)\xi (s)}},  \label{s}
\end{equation}
\begin{equation}
\xi (s)=\left\{ \begin{array}{ll}
             2              & \mbox{if $s \le s_{1}$}\\*[6pt]
             1+\ln s/\ln s_{1} & \mbox{if $s_{1}< s < 1$},\\*[6pt]
             1              & \mbox{if $s \ge 1$}
\end{array}   
\right.
\label{e}
\end{equation}
($\sqrt{s_{1}}= Z^{1/3}/184$) and
\begin{eqnarray}
G(s)&=&\displaystyle 48s^{2}\left\{\frac{\pi}{4}+
          {\rm Im}\,\Psi\!\left(s+\frac{1}{2}+is\right)\right\},
          \label{G1}\\*[6pt]
\Phi (s)&=&\displaystyle 12s^{2}\left\{{\rm Im}\,[\Psi(s+is)+\Psi(s+1+is)]-
\frac{1}{2}\right\},  \label{Phi1}  
\end{eqnarray}
where $\Psi$ represents the logarithmic derivative of the complex
$\Gamma$ function.
The functions $G(s)$ and $\Phi (s)$ are plotted in figure 
\ref {fig:Gphi}. Both $G(s)$ and $\Phi(s)$
belong to the interval $[0,1]$ for all $s \geq 0$ and their limits
for $s$ going to zero (infinity) are 0 (1). Therefore, when $s \ll 1$ the
suppression is important (the cross section
is reduced, $S \ll 1$), while for $s \gg 1$ there is no suppression
($S \cong 1$) and the Migdal cross section reproduces the main terms of
the Bethe-Heitler equation \cite{BH} if complete screening is considered, 
namely
\begin{equation}
\frac{d\sigma _{0}}{dk}=\frac{4\alpha r_{e}^{2}}{3k}%
\{y^{2}+2[1+(1-y)^{2}]\}Z^{2}\ln \left(\frac{184}{Z^{\frac{1}{3}}}\right).
\label{BH}
\end{equation}
It is easy to see, for example, that if the medium is air\footnote{For
altitudes up to $90$ $\mathrm{km}$ above sea level, the air is a   
mixture of $78.09 \%$ $N_{2}$, $20.95 \%$ $O_{2}$ and $0.96 \%$ other
gases \cite{usatmCRC} which can be adequately modeled as an
homogeneous substance with atomic charge and mass numbers
$Z_{\mathrm{eff}}=7.3$ and $A_{\mathrm{eff}}=2 \times
Z_{\mathrm{eff}}$, respectively.} at the altitude $h$ which
corresponds to a vertical depth $X_{v}=1000$
$\mathrm{g}/\mathrm{cm}^{2}$ ($X_{v}(h)=\int_{h}^{\infty} \rho(z)dz$
\cite{Aires}, $X_{v}(0)\cong 1000$ $\mathrm{g}/\mathrm{cm}^{2}$,
$X_{v}(100\;\mathrm{km})\cong 0$) and for electron energies of 100 TeV, 
the screened\footnote{The screening effect of the outer
electrons has been calculated by Bethe and Heitler on the basis of the
Fermi-Thomas model of the atom. The influence of the screening on a
radiation process is determined by the quantity 
$\gamma=100(m/E)[y/(1-y)]Z^{-1/3}$. For $\gamma \gg1$ screening can be
practically neglected while one has complete screening for $\gamma \cong
0$ \cite{BH,rossi}.} Bethe-Heitler equation and the no suppression limit
(\ref {BH}) of Migdal theory agree within the $2.5 \% $ relative error
level. In concordance with equation (\ref {rel})
the region of photon spectrum where the LPM suppression is significant
grows when the primary energy is enlarged, and affects virtually the
complete range $0< y < 1$ when the energy of the electron is $10^{18}$
eV (in this case $E_{\rm LPM} \cong 2.8\times 10^{17}$ eV $=280$
PeV).

The plots in figure \ref {fig:Mbress} illustrate the dependence of the
LPM effect with the density of the medium. In this figure the
probability
\begin{equation}
\frac{dP}{dk}=\eta X_{o}\frac{d\sigma}{dk}, \label{PP}
\end{equation}
is plotted versus $y$ for several electron
energies. The medium is air at two representative altitudes: (a)
Vertical atmospheric depth $X_{v}=1000$ $\mathrm{g}/\mathrm{cm}^{2}$,
which corresponds approximately to an altitude of 300 m.a.s.l
($\rho=1.19$ $\mathrm{kg}/\mathrm{m}^{3}$, $E_{\rm LPM} \cong 280$
PeV); (b) $X_{v}=50$ $\mathrm{g}/\mathrm{cm}^{2}$, which corresponds
approximately to 20,000 m.a.s.l ($\rho=78$
$\mathrm{g}/\mathrm{m}^{3}$, $E_{\rm LPM} \cong 5.78$ EeV).  The plots
in figure \ref{fig:Mbress}a show that an important suppression takes
place at all the energies considered and becomes severe for the
$10^{20}$ eV case.  On the other hand, the LPM
effect does not seriously affect the bremsstrahlung probabilities for
$10^{18}$ eV electrons at $X_{v}=50$ $\mathrm{g}/\mathrm{cm}^{2}$ as
shown in figure \ref{fig:Mbress}b. In this case the critical energy
where the effect becomes significant is placed around $10^{19}$ eV
since $E_{\rm LPM}$ is some 20 times larger than the corresponding
value for figure \ref{fig:Mbress}a.

When the effect of the polarization of the medium is taken into account, a
significant alteration in the bremsstrahlung formula for soft photons
appears. The interaction of a photon with the atomic electrons produces 
another kind of suppression of the cross section that is
called {\em dielectric suppression} \cite{Migdal,TM}. If we take into
consideration that the dielectric constant of the medium is different from
one, i.e., $\varepsilon
=1-(\hbar \omega_{p})^{2}/k^{2}$, where $\omega_{p}$ is the well-known 
plasma frequency ($ \omega_{p}^{2}=4\pi Z e^{2} \eta /m$; for air  in
normal conditions $\hbar\omega_{p}=0.72$ eV), then the longitudinal
momentum transferred to a nucleus changes from equation (\ref {q}) to
\begin{equation}
q_{\parallel}=\frac{km^{2}c^{3}}{2E(E-k)}
+\frac{(\hbar\omega_{p})^2}{2ck}. \label{q3}
\end{equation}
Consequently, when the second term in the last equation is comparable with 
the first, the dielectric suppression becomes important. This happens
when the energy of the photon is much lower than a critical energy given
by $k_{\rm crit} = \hbar \omega_{\rm crit}$ with
\begin{equation}
\omega_{\rm crit}= \omega_{p}\sqrt{\frac{E(E-k)}{m^{2}c^{4}}}\cong
 \omega_{p}\frac{E}{mc^{2}} \label{kcrit}
\end{equation}

It is worth to notice that the correction due to the dielectric
constant is certainly negligible for the propagation phenomenon. On the
contrary, in our case of the emission process, the effect is measurable
because here the scale fixing parameter is the critical frequency given
by equation (\ref {kcrit}), and not the plasma frequency.
Notice also that this critical frequency is exactly the plasma
frequency when measured in the electron rest frame. If we define now
\begin{equation}
y_{\rm diel}=\frac{\hbar\omega_{p}}{mc^{2}}, \label{ydiel}
\end{equation}
equations (\ref {q3}) and (\ref {kcrit}) tell us that the dielectric
suppression takes place when $y \ll y_{\rm diel}$.

The Migdal approach takes into account this dielectric effect.
Since dielectric suppression occurs for $y \ll y_{\rm diel}\ll 1$, the term in
$G(s)$ can be neglected. In that case, the cross section becomes
\cite{Migdal}
\begin{equation}
\frac{d\sigma _{\rm LPM}}{dk}=\frac{16\alpha r_{e}^{2}\xi (s)}{3k \delta}
\Phi (s\delta) Z^{2}\ln \left(\frac{184}{Z^{\frac{1}{3}}}\right)
, \qquad
\delta=1+\left(\frac{y_{\rm diel}}{y}\right)^{2}.
\label{Blpm2}
\end{equation}

Figure \ref {fig:b+d} illustrates the influence of the dielectric
suppression on the bremsstrahlung cross
section (\ref {Blpm2}) where the probability for bremsstrahlung is
plotted against $y$ in the case of $E=10^{18}$ eV, and for an altitude
of 300 m.a.s.l. The emission
probability is suppressed for $y \ll y_{\rm diel} \cong
1.4\times10^{-6}$. This corresponds to photon energies $k \ll 1$ TeV.
\subsection{Pair production}
The pair production process can be treated similarly as
bremsstrahlung.  The corresponding expression for the cross section of
this interaction, where a photon of energy $k$ produces a pair
$e^{+}e^{-}$ of energies $E$ and $k-E$ respectively, can be obtained
immediately from the bremsstrahlung formula \cite{Klein,Migdal}. In
this case the cross section reads
\begin{equation}
\frac{d\sigma _{\rm LPM}(\gamma \rightarrow e^{+}e^{-})}{dE}=\frac{4\alpha
r_{e}^{2}\xi (\widetilde{s})}{3k}\{G(\widetilde{s})+2[u+(1-u)^{2}]\Phi (%
\widetilde{s})\}Z^{2}\ln \left(\frac{184}{Z^{\frac{1}{3}}}\right)
\label{Plpm}
\end{equation}
where
\begin{equation}
u=\frac{E}{k}  \label{u}
\end{equation}
and
\begin{equation}
\widetilde{s}=\sqrt{\frac{kE_{\rm LPM}}{8E(k-E)\xi (\widetilde{s})}}
\label{s-}
\end{equation}
Clearly the dielectric suppression on the cross section of
the electron-positron pair production can always be neglected. This
follows from the fact that the polarization of the medium influences only
the soft photons whose energies are much smaller than that of the
electrons \cite{TM}.

Figure \ref {fig:Mpares} shows the normalized probability of pair
production at atmospheric depths of 1000 $\mathrm{g}/\mathrm{cm}^{2}$
(a) and 50 $\mathrm{g}/\mathrm{cm}^{2}$ (b) for different energies of
the primary photon. The production probabilities are progressively
suppressed when the primary energy rises. From figure \ref {fig:Mpares},
it is also evident that symmetric processes ($u \sim 0.5$) are the most
affected by the LPM suppression.
\subsection{Comparison with other approaches to the LPM effect}
We should remember that the Migdal approach does not include all the
corrections that should in principle be included. In fact, the inelastic
form factor that accounts for inelastic interactions with the atomic
electrons, and the term that accounts for the Coulomb corrections because
the interaction takes place with the electron in the Coulomb field of the
nucleus \cite{Tsai,Perl} are not taken into account. However, Migdal
formulation proves to work sufficiently well when its results are compared
with experimental data \cite{SLAC}.

Blankenbecler and Drell \cite{BD,B,B97} have computed, in an
alternative way, the magnitude of the LPM suppression. They have used
the eikonal formalism, standard in the study of scattering from an
extended target.  This approach leads to a physically clear
quantum-mechanical treatment of multiple scattering and from that to
the derivation of the LPM suppression of soft photon radiation from
high energy electrons in matter. The approach includes as limiting
cases the Bethe-Heitler \cite{BH} radiation from a charged particle
scattering against an isolated atom, relevant for a thin target and,
in the opposite limit, the LPM effect which suppresses the radiation
in a thick target. Their result for thick targets reads
\begin{equation}
k\frac{dP}{dk}=\frac{2\alpha y m^{2} c^{2}\ell}{\pi E\hbar }(I_{1}-I_{2})
\label{eikon}
\end{equation}
where $\ell$ is the target thickness and  
\begin{eqnarray}
I_{1}&=&w \int_{0}^{\infty}\frac{2+3r(\sqrt{1+4w 
x}-1)}{1+4wx-\sqrt{1+4wx}} \sin x\> dx,      \label{I1} \\*[6pt]
I_{2}&=&\int_{0}^{\infty}\frac{\sin x}{x}\>dx = \frac{\pi}{2}   \label{I2}
\end{eqnarray}
where
\begin{eqnarray}
r&=&\frac{1+(1-y)^{2}}{2(1-y)} ,\label{r} \\*[5pt]
w&=&\frac{E}{6E_{\rm LPM}}\left(\frac{1-y}{y}\right)\frac{\ell}{X_{o}}
\label{w}
\end{eqnarray}
An additional corrective term due to the phase-amplitude correlations
has been calculated in reference \cite{B97}, and must be added to the
probability of equation (\ref{eikon}). In the case of a finite thick
target, this term can be expressed as follows
\begin{equation}                                  \label{eikonc}
\left(k\frac{dP}{dk}\right)_{\!\!c} = \frac{3\alpha (1-y) r}{16 \pi w}
\int_0^{b_l} \left(\frac{2wb_l+1}{z} - 1\right) (z-1)^2 \cos x\> dx,
\end{equation}
with $b_l=\ell /\ell_{f_0}$ and $z=\sqrt{1 + 4 w x}$.

The transition to the LPM regime occurs for $w \cong 1$ and the
extreme LPM limit for $w\gg 1$ in concordance with the relation
(\ref{rel}).

Migdal theory contains a number of approximations that are
not very transparent on physical grounds \cite{BD}.  On the other
hand, the Blankenbecler and Drell formulation is theoretically more
robust in its principles, but does not include dielectric
suppression. Migdal results are simpler to treat numerically and the
differences between the two approaches are small in the case of
infinitely thick media. This shows up clearly in
figure \ref {fig:compaME} where both predictions and their relative
differences are plotted for representative energies.

To make these plots we have evaluated numerically the integrals of
equations (\ref{I1}) and (\ref{eikonc}). Our results can be considered
as a generalization of the analysis presented in reference \cite{BD}
which corresponds to the limiting cases of negligible ($w\ll 1$) or
severe ($w\gg 1$) suppression.

In reference \cite{Klein}, other approaches to the LPM effect
\cite{BK,Saka,BDMS} are reviewed; and similarly as in the case of the
Blankenbecler and Drell formulation their results are not in
contradiction with Migdal theory. It is worth noticing that the main
differences between such different approaches appear when considering
finite thickness slabs, which is not the case of interest in our work,
devoted to the study of the LPM effect in an infinite dilute medium
like the Earth's atmosphere.

\section{Practical Implementation}
\label{PI}

During the computer simulation of a ultra-high energy air shower, the
processes of emission of bremsstrahlung photons and $e^{+}e^{-}$ pair
production need to be calculated some millions of times. This requires
that the algorithms used in such calculations should be fast enough to
let the simulations complete in a moderate amount of time.  The Migdal
formulation for bremsstrahlung and pair production can be implemented
fulfilling those requirements \cite{mocca} and, as discussed in the
previous section, its results are in acceptable agreement with other
alternative approaches.  To fix ideas, we shall illustrate the practical
implementation in the case of bremsstrahlung; the algorithm for pair
production is completely similar.

The probability of equation (\ref {PP}) can be put in the form
\begin{equation}
\frac{dP}{dk}=\frac{dP_{0}}{dk}F(E,k) \label{PP2}
\end{equation}
where
\begin{equation}
\frac{dP_{0}}{dk}=\frac{1}{3k} \label{PP3}
\end{equation}
and
\begin{equation}
F(E,k)=\xi(s) \{ y^{2}G(s)+2[1+(1-y)^{2}]\Phi (s) \} \label{F}
\end{equation}
$s$, $\xi(s)$, $G(s)$ and $\Phi(s)$ are defined in equations
(\ref{s}-\ref{Phi1}).  Therefore it is possible to simulate the
bremsstrahlung processes in two steps \cite{Aires,mocca}.  (i) Rough
approach using the probability (\ref {PP3}), (ii) Correction to (i)
using the rejection approval algorithm to give the exact distribution
(\ref {PP2}). The correction factor is the function $F(E,k)$
adequately normalized.  Step (i) relates with the normal,
Bethe-Heitler bremsstrahlung and can be processed straightforwardly.
To complete step (ii) it is necessary to evaluate $E_{\rm LPM}$, $s$,
$\xi(s)$, $G(s)$ and $\Phi(s)$. $E_{\rm LPM}$ can be evaluated
directly from equation (\ref{Elpm2}) and requires the estimation of
the local air density $\rho = -dX_v/dh$. $s$ and $\xi(s)$ can be
conveniently calculated by means of repeated iterations of equations
(\ref {s}) and (\ref {e}) starting with the initial value
$\xi(s)=1$. This iterative process normally converges in a single
step.  Equations (\ref{G1}) and (\ref{Phi1}) allow to obtain accurate
estimations for $G(s)$ and $\Phi(s)$ using the standard algorithms to
evaluate the complex $\Psi$ function \cite{NumericalRecipe}, but are
not suitable for production procedures which must be fast. In this
case it is better to represent $G(s)$ and $\Phi(s)$ by means of
rational functions:
\begin{eqnarray}
G(s)&=& \frac{12 \pi
s^{2}+a_{3}s^{3}+a_{4}s^{4}}{1+b_{1}s+b_{2}s^{2}+b_{3}s^{3}+a_{4}s^{4}},
\label{G2} \\*[8mm]
\Phi(s)&=& \frac{6s+c_{2}s^{2}+c_{3}s^{3}}{1+d_1 s+d_2 s^2 +c_3 s^3}
 \label{Phi2}
\end{eqnarray}
where the coefficients are adjusted to minimize the maximum relative
error for $s>0$. Notice that equations (\ref {G2}) and (\ref {Phi2})
give exact results for both $s=0$ and $s \rightarrow \infty$
\cite{Migdal}. The fitted coefficients are listed in table 1; the
maximum relative error for $G(s)$ ($\Phi(s)$) is 0.15\% (0.12\%).

The dielectric suppression can be easily implemented as a scaling
factor to $s$, as it clearly follows from equation (\ref{Blpm2}). In
AIRES this scaling factor correction is applied when $y_{\rm diel}/y >
1/2$, i.e., the dielectric suppression is neglected in all cases where
$y_{\rm diel}/y< 1/2$. $y_{\rm diel}$ is evaluated at each case taking
into account the altitude of the particle.
\section{Simulations}
\label{sectsim}

We have investigated the modifications in the bremsstrahlung and
pair production cross sections due to the LPM effect, for individual
processes in different conditions. In this section we are going to
analyze the influence of the effect in the development of air showers
initiated by ultra-high energy astroparticles.

There are two characteristics that must be taken into account when
analyzing the shower development:
\begin{enumerate}
\item {\em The atmosphere is inhomogeneous,\/} the density of the air
diminishes six orders of magnitude when the altitude goes from sea
level to 100 km \cite{usatmCRC}, and therefore the characteristic
energy $E_{\mathrm{LPM}}$ of equation (\ref {Elpm2}) varies
accordingly. As a result, the suppression in the corresponding cross
sections depends strongly on the altitude where they occur. In figure
\ref {fig:Elpm} the energy $E_{\mathrm{LPM}}$ is plotted against the
vertical atmospheric depth, for $X_v$ in the range 0.1
$\mathrm{g}/\mathrm{cm}^2$ (66 km above sea level) to 1000
$\mathrm{g}/\mathrm{cm}^2$ (about sea level). To illustrate the
meaning of this plot let us consider a ultra-high energy
electromagnetic process with a primary energy of, say, $E=10^{20}$ eV
$= 100$ EeV: If the primary particle is located at an altitude of 100
$\mathrm{g}/\mathrm{cm}^2$ the process will be strongly suppressed
whereas if it is located at $X_v<1$ $\mathrm{g}/\mathrm{cm}^2$ the
suppression is negligible. In a similar way, the dielectric
suppression will depend on the altitude since $y_{\rm diel}$ depends
on the air density. However, this parameter does not change with $X_v$
so dramatically as $E_{\rm LPM}$ (compare figures \ref{fig:Elpm} and
\ref{fig:ydiel}).
\item {\em The number of particles in the ultra-high energy showers is
very large\/} (about $10^{11}$ for a primary energy of $10^{20}$ eV),
and therefore the influence of the LPM effect and/or dielectric
suppression on {\em global\/} observables may not be visible if the
fraction of affected particles is not large enough.
\end{enumerate}

To estimate how the LPM effect can modify the shower development when
the preceding features are taken into account, we have run a set of
computer simulations using the AIRES program \cite{Aires} to determine
what is the average fraction of electromagnetic particles (gammas,
electrons and positrons) significantly affected by the LPM
suppression.  We have performed simulations using $3\times 10^{20}$ eV
gamma or proton primaries. The particles were injected at the top of
the atmosphere ($100$ km above sea level) with vertical incidence.

At various predetermined altitudes (observing levels), the fraction,
$f_>$, of particles whose energies are greater than $E_{\rm LPM}/4$
were recorded for each simulated shower. This fraction is taken equal
to 1 before the first interaction takes place.\footnote{In the case of
$3\times 10^{20}$ eV gamma showers (LPM effect taken into account),
for example, the mean value of the first interaction depth is
approximately 63 g/cm$^2$ with a standard deviation of 80
g/cm$^2$. These figures become respectively 46 g/cm$^2$ and 45
g/cm$^2$ when the showers are simulated without considering the LPM
effect.}  The data related to every one of the defined levels were
then averaged. In figure \ref {fig:frac} the average fractions,
$\langle f_{>}\rangle$, are plotted versus the vertical depth. The
primaries are gammas. The results coming from simulations where the
LPM effect is taken into account are plotted (full line) together with
similar data obtained without evaluating such effect (dashed line).
We can see that in both cases the relative number of particles
suffering the LPM effect diminishes when $X_v$ grows, reaching almost
zero for points that are always above $500$
$\mathrm{g}/\mathrm{cm}^2$.

Of course, the plots represented in figure \ref {fig:frac} must be
regarded just as qualitative indicators.  Since each shower starts at
a different altitude, the fraction of particles capable of undergoing
the LPM effect will vary accordingly: If a shower starts very high in
the atmosphere, at $X_v<5$ $\mathrm{g}/\mathrm{cm}^2$ for example, it
is likely that the LPM effect will not significantly affect its
development since the first interactions will take place in a region
where $E_{\mathrm{LPM}}$ is very high (see figure \ref {fig:Elpm}). On
the other hand, a penetrating primary particle starting the shower at,
say, $X_v>100$ $\mathrm{g}/\mathrm{cm}^2$ will surely show a
significant modification in its longitudinal development due to the
LPM suppression.

It is worthwhile mentioning that the influence of the LPM effect on
the shower development depends also on the inclination of the shower:
For large zenith angles it is more probable that the showers will
start at points where the density of the air is very low, and so a
smaller suppression should be expected in this case.

In the case of proton primaries, the electromagnetic shower starts
after the first hadronic interactions (mainly after $\pi^0$ decays)
and so the energies involved are smaller than the primary energy. For
this reason, $\langle f_>\rangle$ vanishes faster than in the case of
gamma primaries. In fact, this fraction is virtually zero for $X_v \ge
100$ $\mathrm{g}/\mathrm{cm}^2$ (for simplicity we have not included
any proton related data in figure \ref{fig:frac}).

The most evident signature of the impact of the LPM effect on the
shower development is the shift on the position of the shower maximum,
$X_{\mathrm{max}}.$\footnote{The shower maximum, $X_{\mathrm{max}}$,
is defined as the atmospheric depth of the point where the total
number of charged particles is maximum. The number of charged
particles at $X_{\mathrm{max}}$ is noted as $N_{\mathrm{max}}$.} This
can be easily seen in figure \ref {fig:longcharg} where the total
number of charged particles for showers initiated by gamma rays is
plotted against the vertical depth. These data were obtained by means
of computer simulations using the program AIRES \cite{Aires}, and
correspond to inclined showers with zenith angle 60 degrees; the
primary particles are injected at the top of the atmosphere and the
ground level is located at a vertical depth of $1000$
$\mathrm{g}/\mathrm{cm}^2$. Two primary energies are considered,
namely, $10^{19}$ eV (figure \ref{fig:longcharg}a) and $3\times
10^{20}$ eV (figure \ref{fig:longcharg}b).

The severe suppression suffered by the first interactions that start
the shower shows up clearly in figure \ref{fig:longcharg}b where the
plot corresponding to the simulations performed taking into account the
LPM effect shows significantly more penetrating showers that develop
deeper in the atmosphere in comparison with the non-LPM ones. The position
of $X_{\mathrm{max}}$ is shifted in about 130 $\mathrm{g}/\mathrm{cm}^2$
(about 260 $\mathrm{g}/\mathrm{cm}^2$ measured along the shower axis)
when the LPM procedures are switched on. The number of particles at
the maximum, $N_{\mathrm{max}}$ is also affected: It is about a 40\%
smaller in the LPM suppressed showers. This is mainly due to the fact that
the electrons and positrons loss more energy in ionization
processes before being able to generate bremsstrahlung photons, and
therefore the shower reaches its maximum generating less particles
than in the non-LPM case.

For lower primary energies the influence of the LPM effect on the
development of the shower is less significant, as shown in figure \ref
{fig:longcharg} a for the case of $10^{19}$ eV gamma showers.
Accordingly with the results of our simulations, the LPM effect does
not seriously modify the shower longitudinal profile (in the case of
gamma primaries) when the primary energy is less than $10^{18}$ eV.

The longitudinal development of the number of muons is also dependent
on the LPM effect as shown in figure \ref {fig:longmuons}.

These plots also show the results of simulations made with the LPM
effect switched on, but without considering the dielectric
suppression. The impact of this last effect is of course less
important but not small enough to justify not considering it, and in
fact it is more important in relative terms than in the case of the
longitudinal development of charged particles (figure
\ref{fig:longcharg}). It is interesting to describe how the number of
muons can be altered when the dielectric suppression of electron
bremsstrahlung photons is taken into account: In the case of electrons
of energy slightly below the primary energy, that is, about $10^{18}$
eV, the dielectric suppression diminishes the probability of emission
of photons with energies below 10 GeV ($y \approx 10^{-8}$), producing
a relative enlargement of the number of events with photon energies
slightly above this limit. Such photons can undergo photonuclear
reactions creating pions which in turn may decay into muons; and it is
clear that if their number is enlarged so will be the number of
secondary muons (compare the full and dotted lines of figure
\ref{fig:longmuons}b).

If the primary particle is not electromagnetic, the influence of the
LPM effect will not depend directly on the primary energy, but on the
energy of the particles that initiate the electromagnetic shower. In
the case of showers with hadronic primaries, protons for example, the
electromagnetic shower is typically started by the gamma rays product
of decays of $\pi^0$ mesons produced after inelastic hadronic
processes that take place when the primary collides against an
atmospheric nucleus. For primary energies larger than $10^{19}$ eV, an
inelastic collision hadron-nucleus produces hundreds of secondaries,
most of them having energies which are 1 to 4 orders of magnitude
smaller than the primary energy. Roughly speaking, this means that the
electromagnetic shower maximum energy is 2 or 3 orders of magnitude
below the energy of the primary, and accordingly with the data
presented in figure \ref {fig:longcharg}, this means that the LPM
effect will not produce serious distortion in the shower development
unless the energy of the primary hadron is well above $10^{21}$ eV.
This agrees with our results presented in figure \ref {fig:longproton}
which correspond to the longitudinal development of $3\times 10^{20}$
eV proton showers: The curves corresponding to both LPM ``on'' and
``off'' cases do not present significant differences.

Another remarkable feature is the increase of the {\em fluctuations\/}
of $X_{\mathrm{max}}$ and $N_{\mathrm{max}}$. In figure \ref
{fig:xmax} (\ref {fig:nmax}) $X_{\mathrm{max}}$ ($N_{\mathrm{max}}$)
and its fluctuations are plotted as functions of the primary energy,
for showers in the range $10^{14}$ -- $10^{21}$ eV. It shows up
clearly that: (i) The well-known proportionalities between
$X_{\mathrm{max}}$ and $\log N_{\mathrm{max}}$ with the logarithm
of the primary energy \cite{Gaisser,SIBYLL2} do not hold exactly for
electromagnetic showers of ultra-high energy when the LPM effect is
taken into account.  However, the mentioned linear relations are valid
for proton showers in the whole range of energies considered. (ii) As
mentioned, the fluctuations of $N_{\mathrm{max}}$ and especially those
of $X_{\mathrm{max}}$ become larger as long as the primary energy is
increased, in comparison with the non-LPM counterparts.  The results
presented in this section correspond to particular cases which are
representative of the behavior that should be expected in other
general cases. Notice that the characteristics of electron initiated
showers are very similar to those corresponding to gamma showers and
the same remark can be done in the case of showers initiated by nuclei
in comparison with proton showers. For that reason we are not
including here any related plots.
\section{Conclusions}
\label{conclu}

We have analyzed exhaustively the LPM effect and the dielectric
suppression from the theoretical point of view and have discussed the
practical implementation of the corresponding algorithms for air
shower simulations.

We have studied the different approaches to the LPM effect and
conclude that the final results of the Migdal theory are best adapted
for the numerical treatment while there are no important differences
with the results coming from other approaches.

By means of numerical simulations using AIRES \cite{Aires}, we have
studied the influence of the LPM effect, including dielectric
suppression, on the longitudinal development of air showers initiated
by ultra-high energy astroparticles. As mentioned previously, AIRES is
capable of calculating many air shower observables in a realistic
environment.

We have analyzed the influence of both suppression mechanisms in a
wide range of primary energies (up to $3\times 10^{20}$ eV), and for
various primary particles, namely, photons, electrons and protons.

The main purpose of our work has been to study the influence of the
LPM effect on the behavior of electromagnetic air showers, for
primaries that interact after reaching the Earth's atmosphere.

Clearly, in the special case of cosmic gamma rays, it would be
necessary to take into account the interaction of the primary with the
geomagnetic field that could take place before entering the atmosphere
\cite{Herber,Vankov,Pierre}. The cosmic photon can create
electron-positron pairs, which in turn may originate additional
magnetic bremsstrahlung photons, and so on. As a result, a {\em
pre-shower\/} takes place before the particles reach the atmosphere,
and consequently the original energy is shared among various
electromagnetic particles. If the number of particles in the
pre-shower is large, the influence of the LPM effect will be
substantially reduced, in comparison with the case of a single
electromagnetic particle. Nevertheless, the conversion probability
strongly depends on the initial conditions (geomagnetic field
strength, angle between the velocity of the particle and the magnetic
field, etc.) \cite{Pierre}, and can be relatively small in certain
circumstances. Consequently, the pre-shower may eventually not exist
or be very small, allowing ultra-high energy electromagnetic particles
to reach the atmosphere, producing showers like those included in our
study.

 We have found that the LPM effect introduces significant
modifications on the development of gamma and electron air showers if
the primary energies are larger than $10^{19}$ eV. The most evident
signature of the effect is the shift in the position of the maximum of
the shower, which moves deeper into the atmosphere with increasing
fluctuations when the primary energy is enlarged. Our conclusion is
that in such cases the effect must be always taken into account in
realistic simulations of ultra-high energy electromagnetic air
showers. It is also important to remark that: (i) The influence of the
dielectric suppression is not as important as the LPM effect, but
large enough to justify having it taken into account in any realistic
simulation. (ii) For showers with large zenith angles, the suppression
that delays the shower growth can be not as large as in the case of
vertical showers, as explained in section \ref{sectsim}.

We have not found any important effect in proton showers with primary
energies up to $10^{21}$ eV. The reason for this, as explained in the
previous section, is that the electromagnetic shower, where the LPM
effect takes place, begins later, when the initial energy is shared
among the secondary particles and the average energy per particle is
then 2-3 orders of magnitude less than the primary energy. Clearly the
same reasoning is valid for nuclei primary cosmic rays.

\section*{Acknowledgments}

We are indebted to C. Hojvat for useful discussions and comments. This
work was partially supported by the Consejo Nacional de
Investigaciones Cient\'{\i}ficas y T\'ecnicas of Argentina (CONICET).

\def\journal#1#2#3#4{{\em #1,\/} {\bf #2}, #3 (#4)}

\clearpage
\def\epsfig#1{\epsfbox{#1}}
\begin{figure}
\begin{displaymath}
\hbox{\epsfig{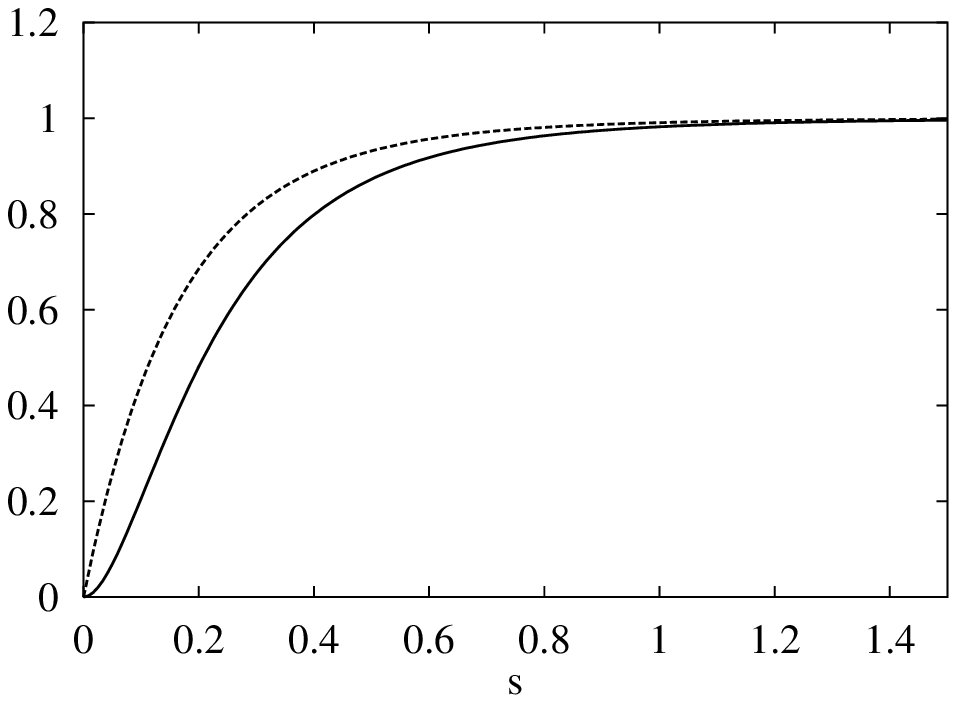}} 
\end{displaymath}
\caption{Functions $G(s)$ (solid line) and\/ $\Phi(s)$ (dashed line),
appearing in Migdal theory. The plots correspond to {\em exact\/}
calculations made by numerical evaluation of equations (\ref{G1}) and
(\ref{Phi1}).}
\label{fig:Gphi}
\end{figure}
\clearpage
\begin{figure}[tb]
\begin{displaymath}
\begin{array}{cc}
\hbox{\epsfxsize=7.4cm\epsfig{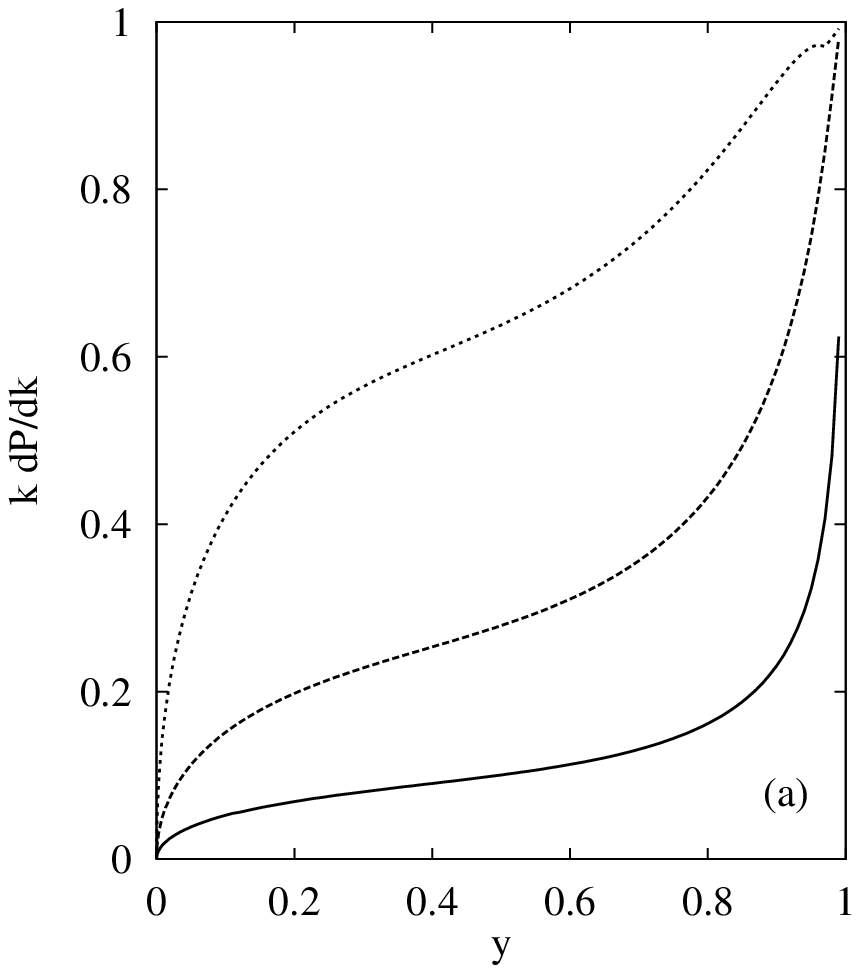}} &
\hbox{\epsfxsize=7.4cm\epsfig{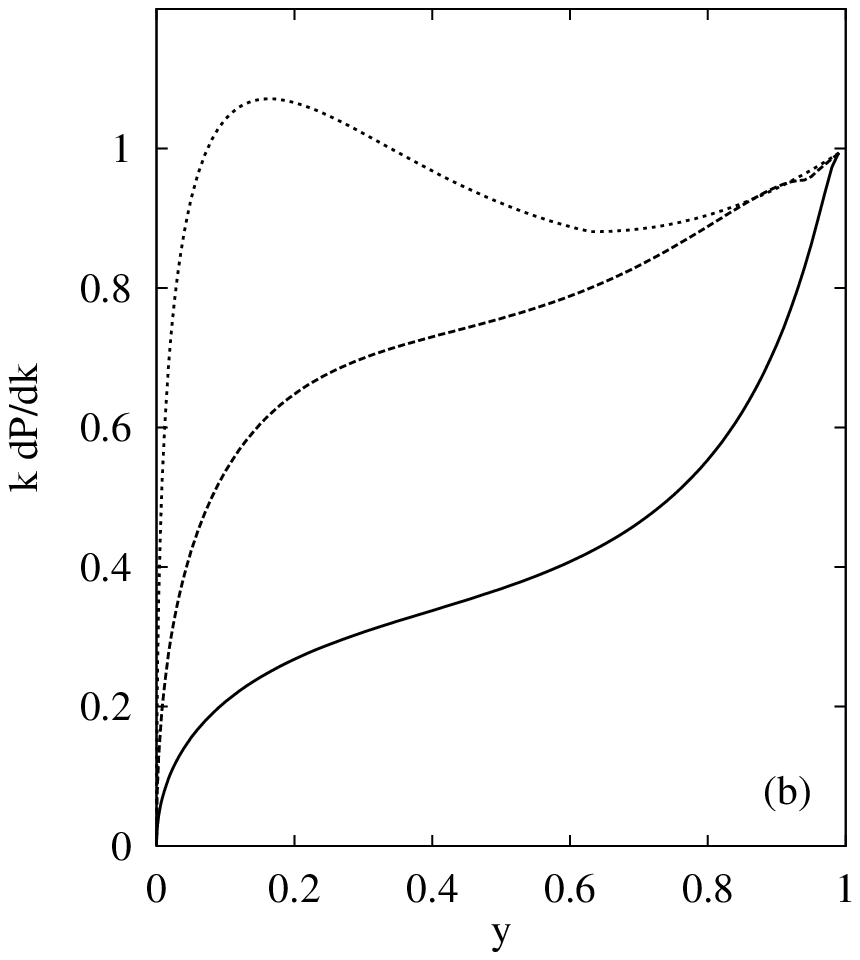}}
\end{array}
\end{displaymath}
\caption{ Bremsstrahlung probability according to Migdal theory
($kdP/dk$). The electron energies are $E=10^{18}$ {\rm eV} (solid
line), $E=10^{19}$ {\rm eV} (dashed line) and $E=10^{20}$ {\rm eV}
(dotted line). The medium is air at the vertical depth of $X_{v}=1000$
$\mathrm{g}/\mathrm{cm}^{2}$ ($\rho=1.19$
$\mathrm{kg}/\mathrm{m}^{3}$) (a), and $X_{v}=50$
$\mathrm{g}/\mathrm{cm}^{2}$ ($\rho=78$ $\mathrm{g}/\mathrm{m}^{3}$)
(b).}
\label{fig:Mbress}
\end{figure}
\begin{figure}
\begin{displaymath}
\hbox{\epsfxsize=10cm\epsfig{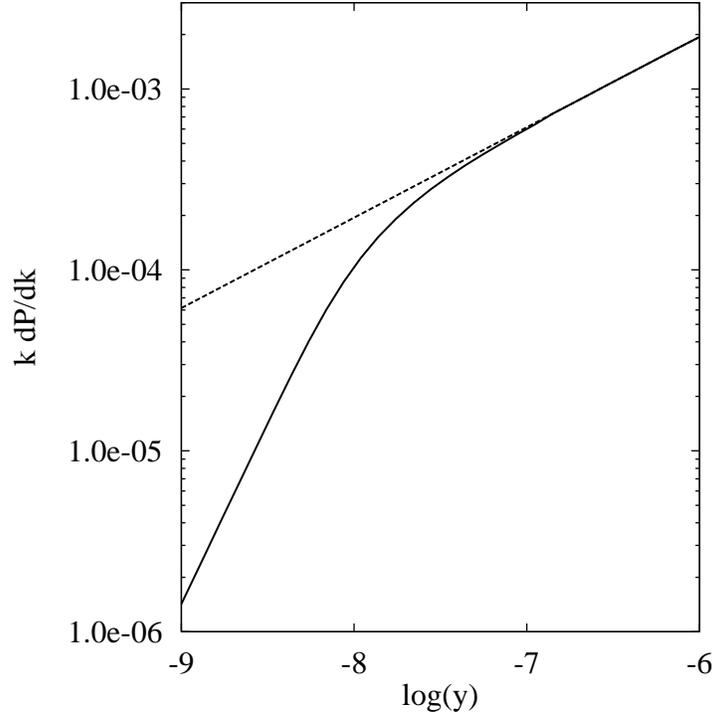}}
\end{displaymath}
\caption{Bremsstrahlung probability according
to Migdal theory ($kdP/dk$). The probabilities taking (solid line) and
not taking (dashed line) into account the influence of dielectric
suppression are plotted versus $\log_{10}y$. The energy of the electron
is $10^{18}$ {\rm eV} and the medium is air at a vertical depth
$X_{v}=1000$ $\mathrm{g}/\mathrm{cm}^{2}$ ($\rho=1.19$
$\mathrm{kg}/\mathrm{m}^{3}$).}
\label{fig:b+d}
\end{figure}
\begin{figure}
\begin{displaymath}
\begin{array}{cc}
\hbox{\epsfxsize=7.4cm\epsfig{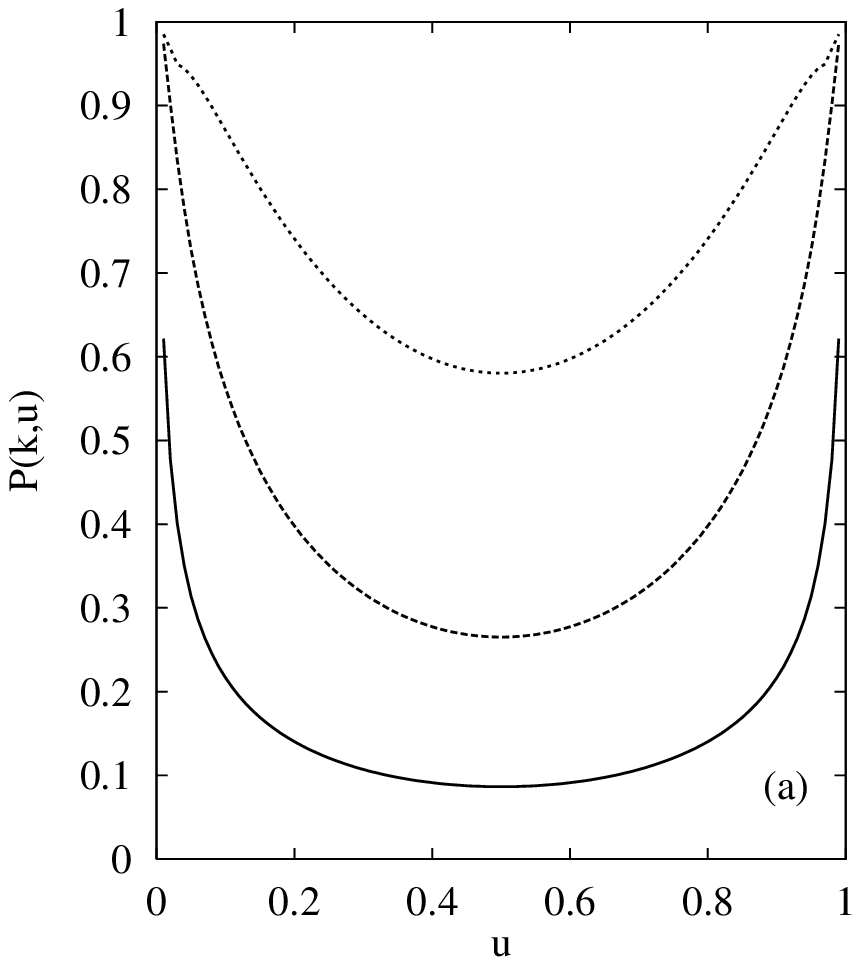}} &
\hbox{\epsfxsize=7.4cm\epsfig{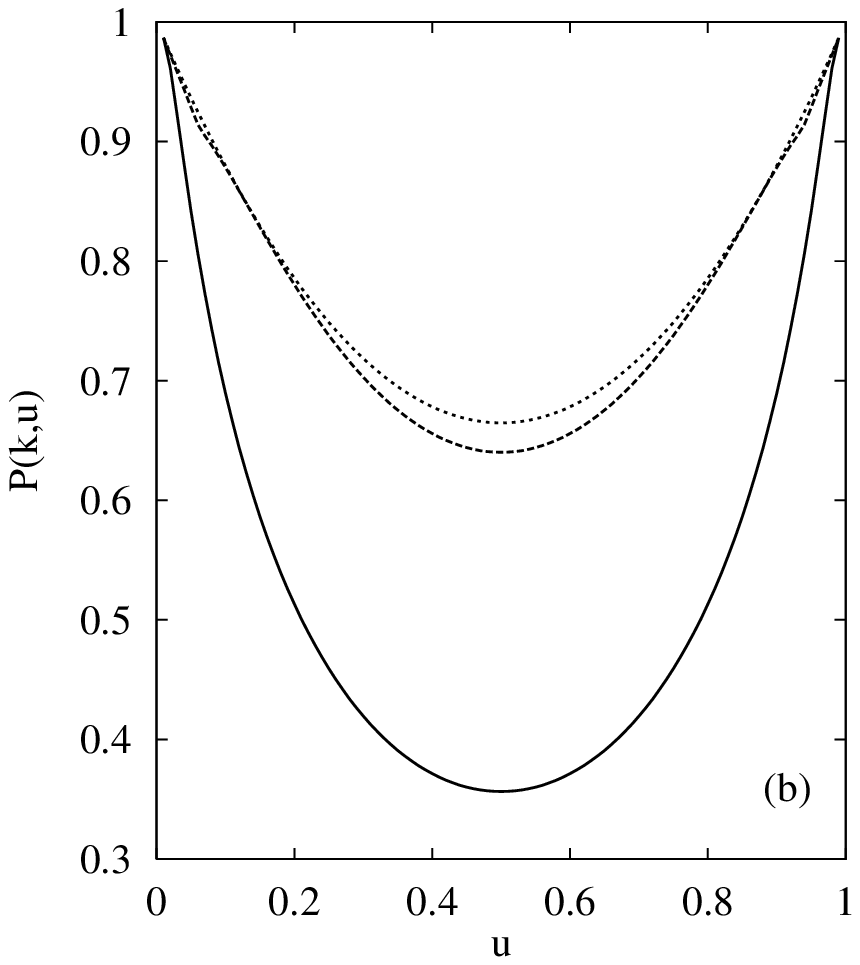}}
\end{array}
\end{displaymath}
\caption{Pair production probability, $P(k,u)$, according to Migdal
theory plotted versus $u$ for different values of the photon energy:
$10^{20}$ {\rm eV} (solid line), $10^{19}$ {\rm eV} (dashed line) and
$10^{18}$ {\rm eV} (dotted line). The medium is air at the vertical
depth of $X_{v}=1000$ $\mathrm{g}/\mathrm{cm}^{2}$ ($\rho=1.19$
$\mathrm{kg}/\mathrm{m}^{3}$) (a), and $X_{v}=50$
$\mathrm{g}/\mathrm{cm}^{2}$ ($\rho=78$ $\mathrm{g}/\mathrm{m}^{3}$)
(b).}
\label{fig:Mpares}
\end{figure}
\begin{figure}
\begin{displaymath}
\begin{array}{cc}
\hbox{\epsfxsize=7.4cm\epsfig{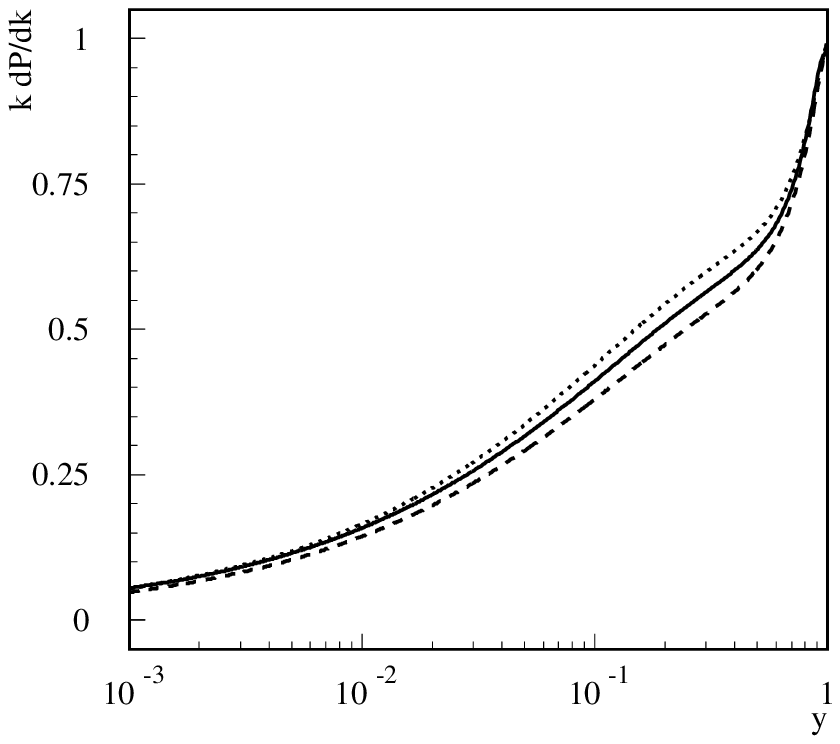}} &
\hbox{\epsfxsize=7.4cm\epsfig{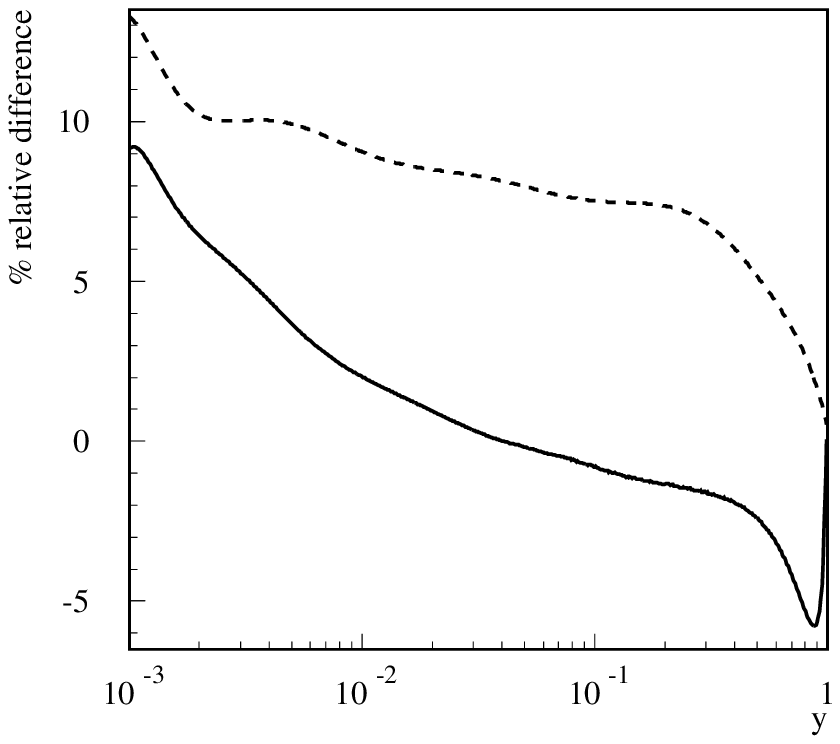}}
\end{array}
\end{displaymath}
\caption{Comparison between Migdal theory and Blankenbecler and Drell
formulation. (a) Absolute probabilities versus $y$. The electron
energy is $10^{18}$ {\rm eV}, and the solid (dashed) line represents
Migdal theory (Blankenbecler and Drell formulation). The dotted line
corresponds to the Blankenbecler and Drell formulation without adding
the phase-amplitude correlation correction (\ref{eikonc}).  (b)
Percent relative differences between both formulations, for electron
energies $10^{18}$ {\rm eV} (dashed line) and $10^{20}$ {\rm eV}
(solid line). In all cases the medium is air and the vertical depth
is $X_{v}=1000$ $\mathrm{g}/\mathrm{cm}^{2}$ ($\rho=1.19$
$\mathrm{kg}/\mathrm{m}^{3}$) ($10^{18}$ eV) and $X_{v}=100$
$\mathrm{g}/\mathrm{cm}^{2}$ ($\rho=0.156$
$\mathrm{kg}/\mathrm{m}^{3}$) ($10^{20}$ eV).}
\label{fig:compaME}
\end{figure}
\begin{figure}
\begin{displaymath}
\hbox{\epsfig{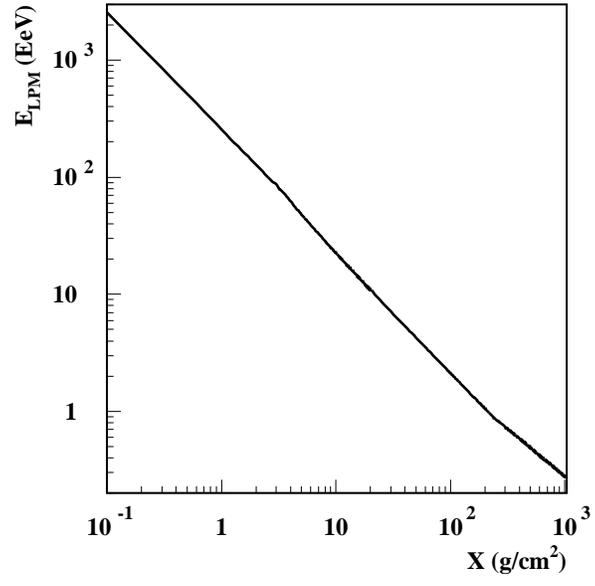}}
\end{displaymath}
\caption{$E_{\rm LPM}$ versus vertical atmospheric depth.}
\label{fig:Elpm}
\end{figure}
\begin{figure}
\begin{displaymath}
\hbox{\epsfig{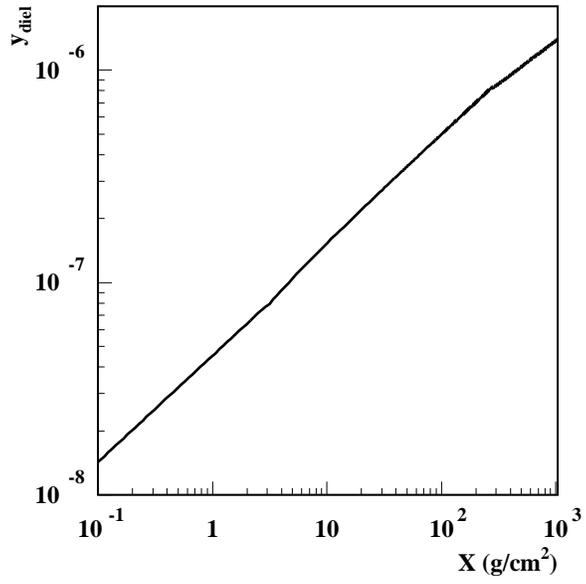}}
\end{displaymath}
\caption{$y_{\rm diel}$ versus vertical atmospheric depth.}
\label{fig:ydiel}
\end{figure}
\begin{figure}
\begin{displaymath}
\hbox{\epsfig{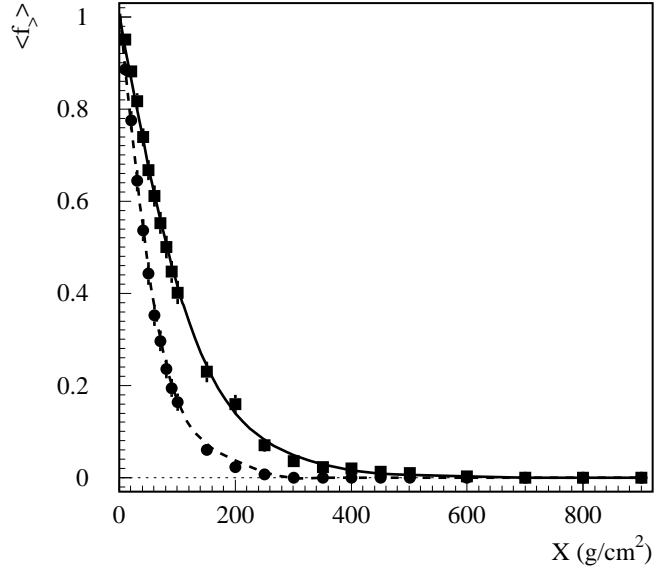}}
\end{displaymath}
\caption{Average fraction of air shower particles with energy larger
than $E_{\rm LPM}/4$, $\langle f_{>}\rangle$, plotted as a function of
the vertical atmospheric depth. The lines are informal fits to
simulation data obtained using AIRES. The solid (dashed) line
correspond to showers initiated by gamma primaries propagated taking
(not taking) into account the LPM effect.  In both cases the primary
energy is\/ $3\times10^{20}$ {\rm eV} and the shower axis is
vertical.}
\label{fig:frac}
\end{figure}
\begin{figure}
\begin{displaymath}
\begin{array}{cc}
\hbox{\epsfxsize=7.4cm\epsfig{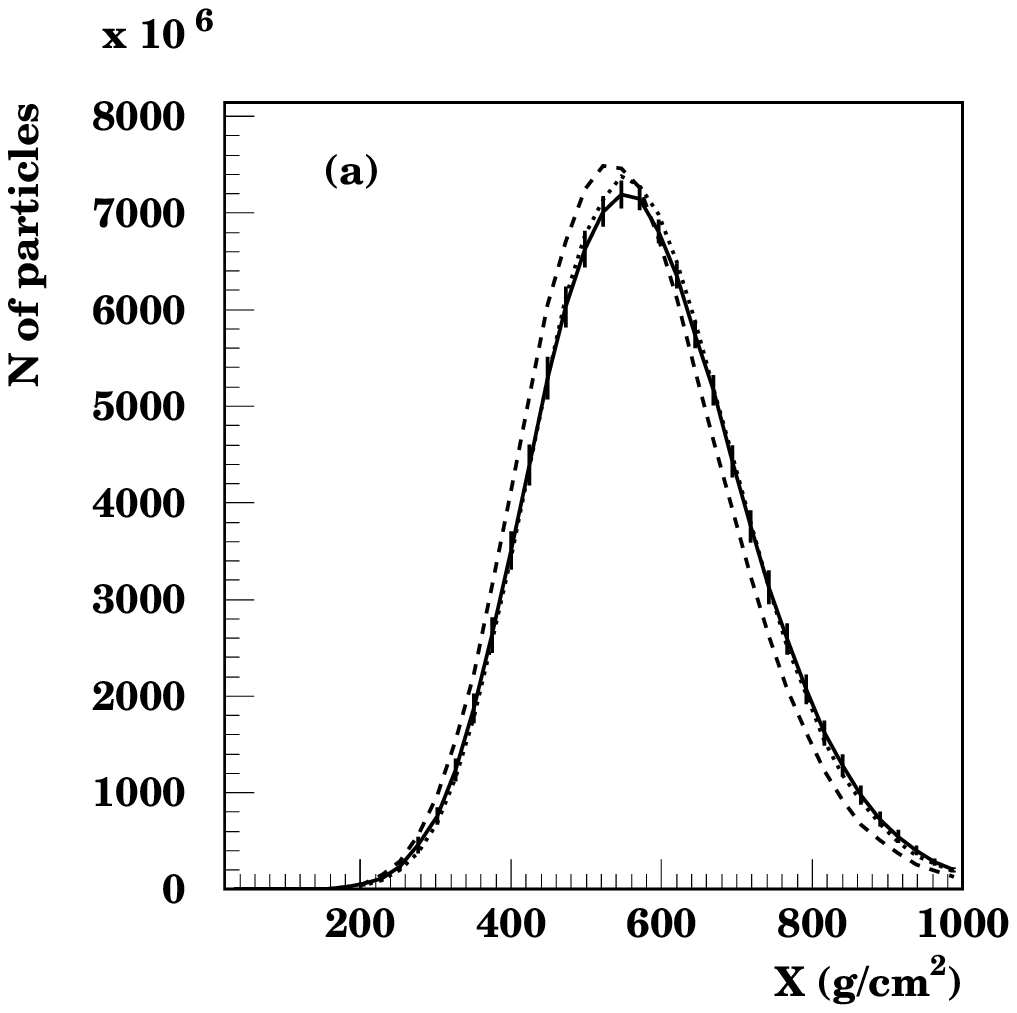}} &
\hbox{\epsfxsize=7.4cm\epsfig{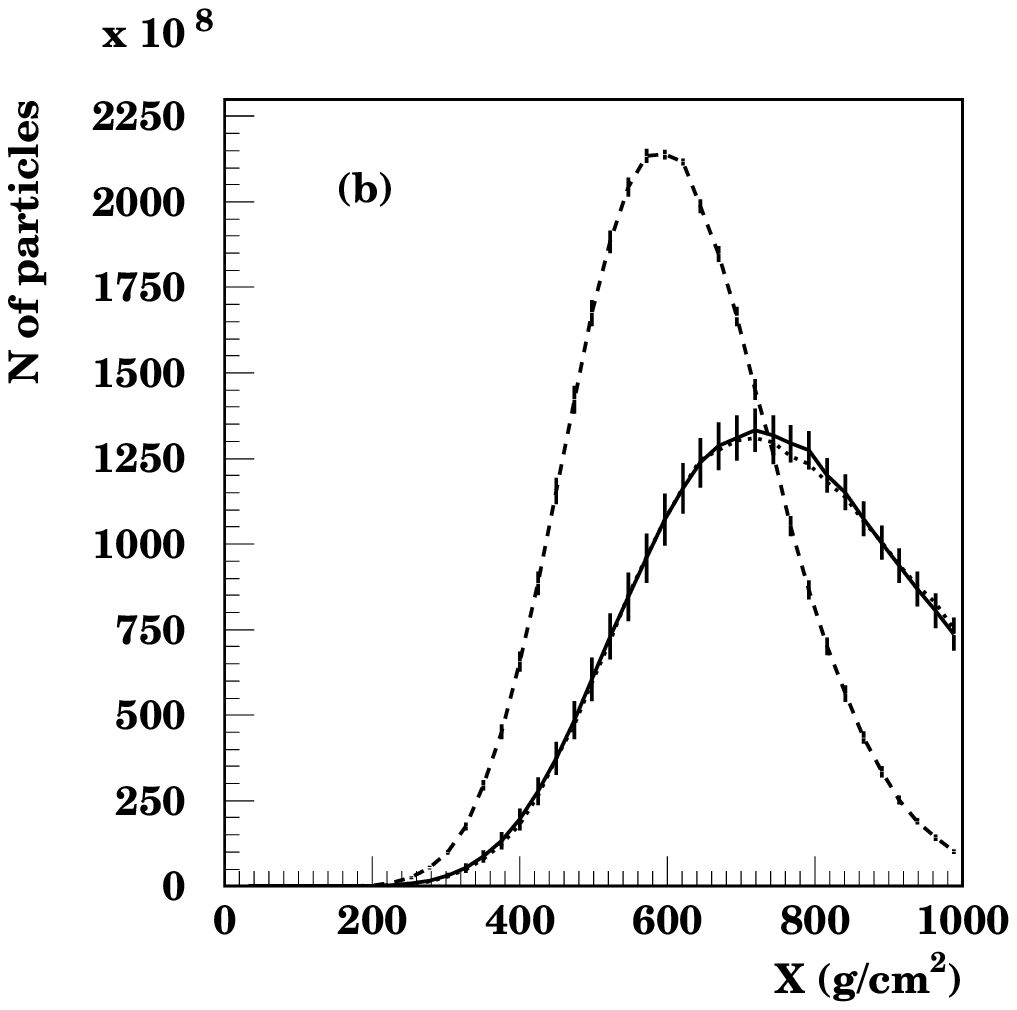}}
\end{array}
\end{displaymath}
\caption{Longitudinal development of charged particles, obtained by
means of Monte Carlo simulations using AIRES, plotted
versus the vertical atmospheric depth. The primary particle is a
photon injected at the top of the atmosphere (100 km above sea level)
with a zenith angle of 60 degrees. The primary energies are $10^{19}$
{\rm eV} (a) and $3\times10^{20}$ {\rm eV} (b). The solid (dashed)
lines correspond to calculations that include (do not include) the LPM
effect. The dotted lines correspond to the LPM case but without the
dielectric suppression. In some cases the error bars (the RMS errors
of the means) were not plotted for clarity; they are, in general,
smaller or comparable to the represented ones.}
\label{fig:longcharg}
\end{figure}
\begin{figure}
\begin{displaymath}
\begin{array}{cc}
\hbox{\epsfxsize=7.4cm\epsfig{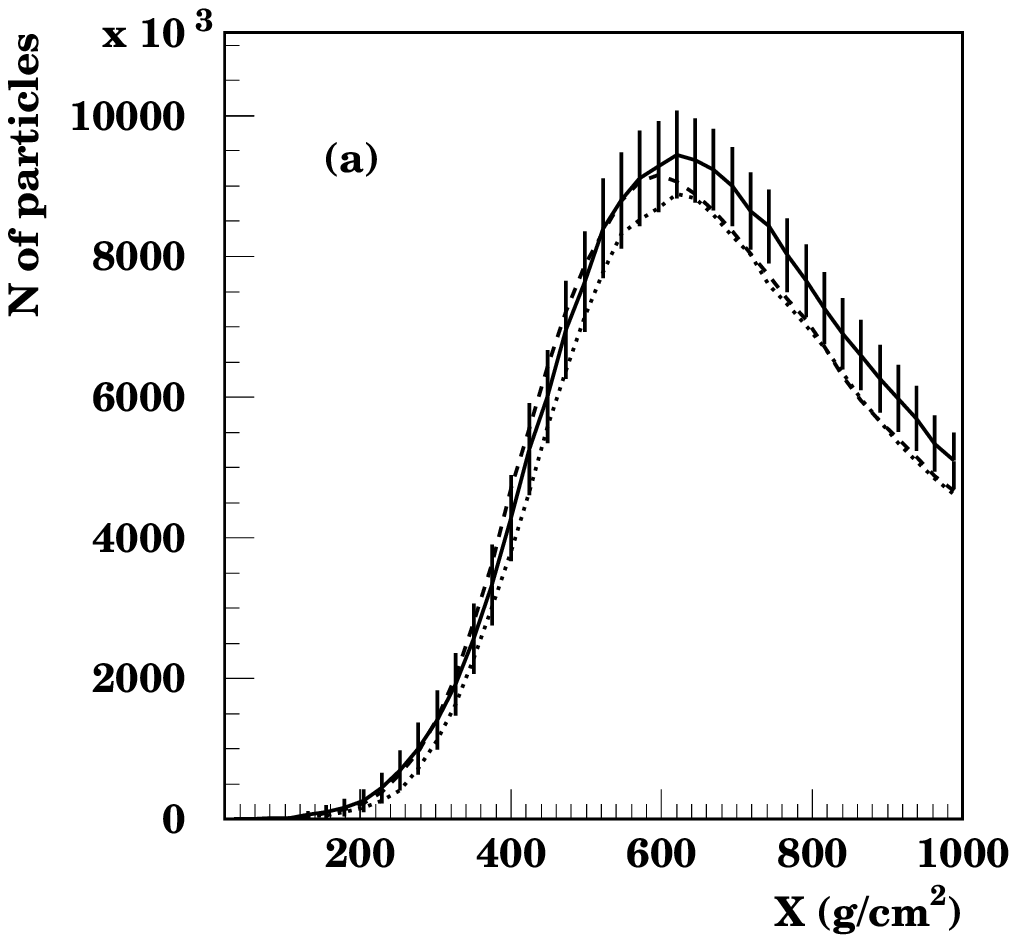}} &
\hbox{\epsfxsize=7.4cm\epsfig{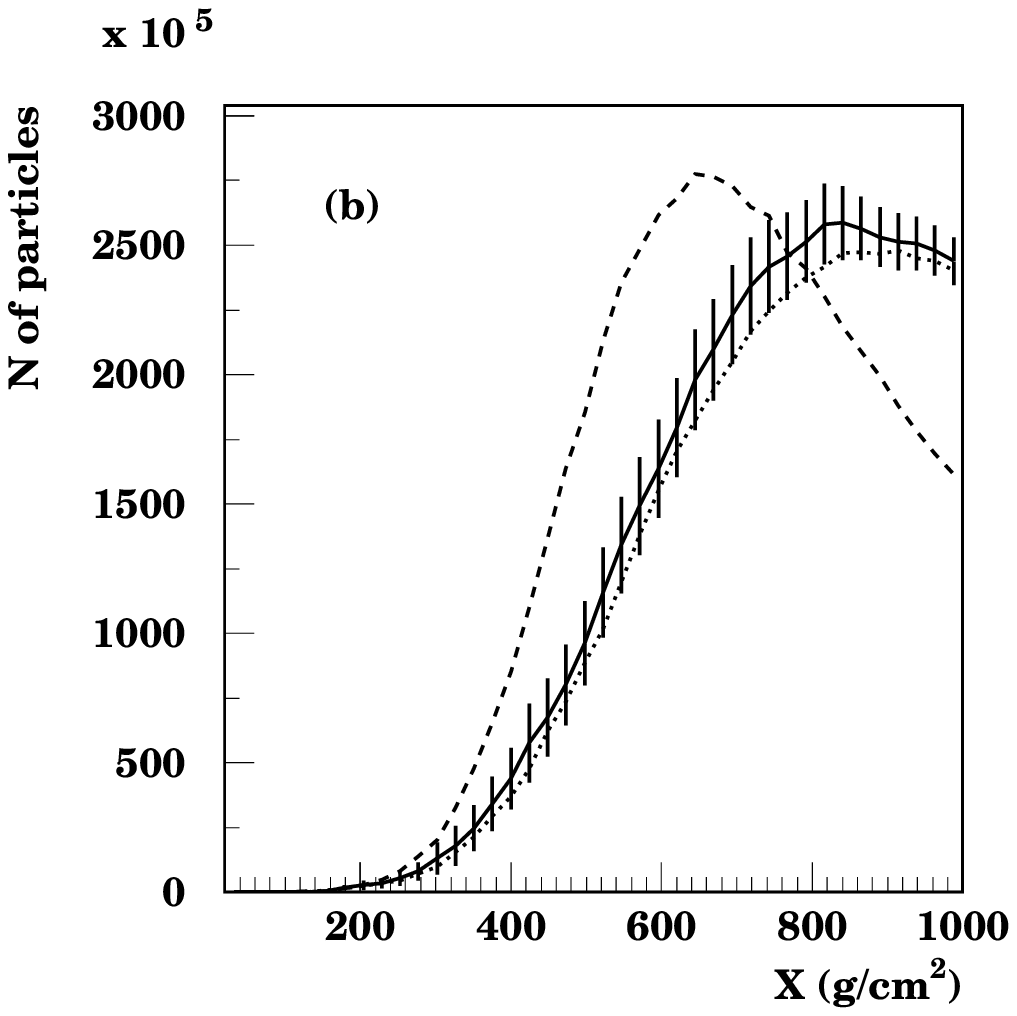}}
\end{array}
\end{displaymath}
\caption{Same as figure \ref{fig:longcharg}, but for the longitudinal
development of the number of muons.}
\label{fig:longmuons}
\end{figure}
\begin{figure}
\begin{displaymath}
\begin{array}{cc}
\hbox{\epsfxsize=7.4cm\epsfig{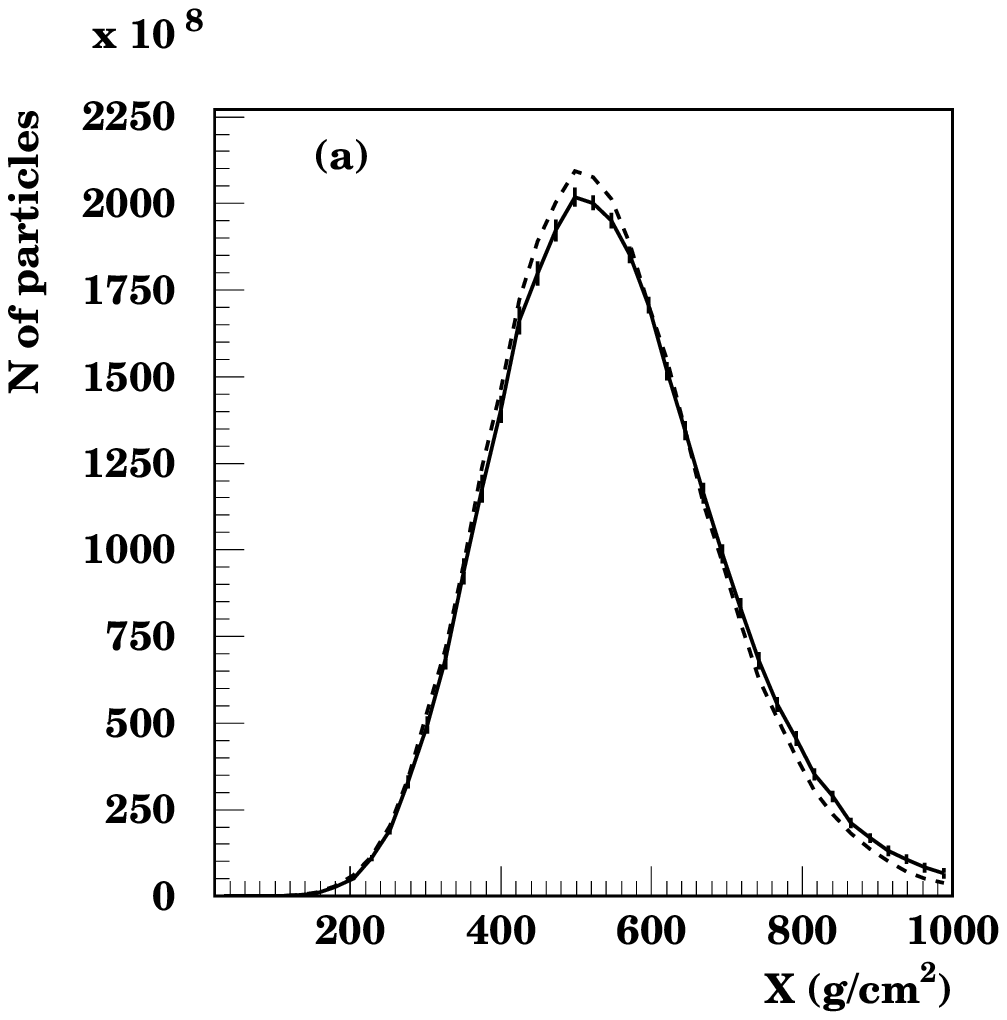}} &
\hbox{\epsfxsize=7.4cm\epsfig{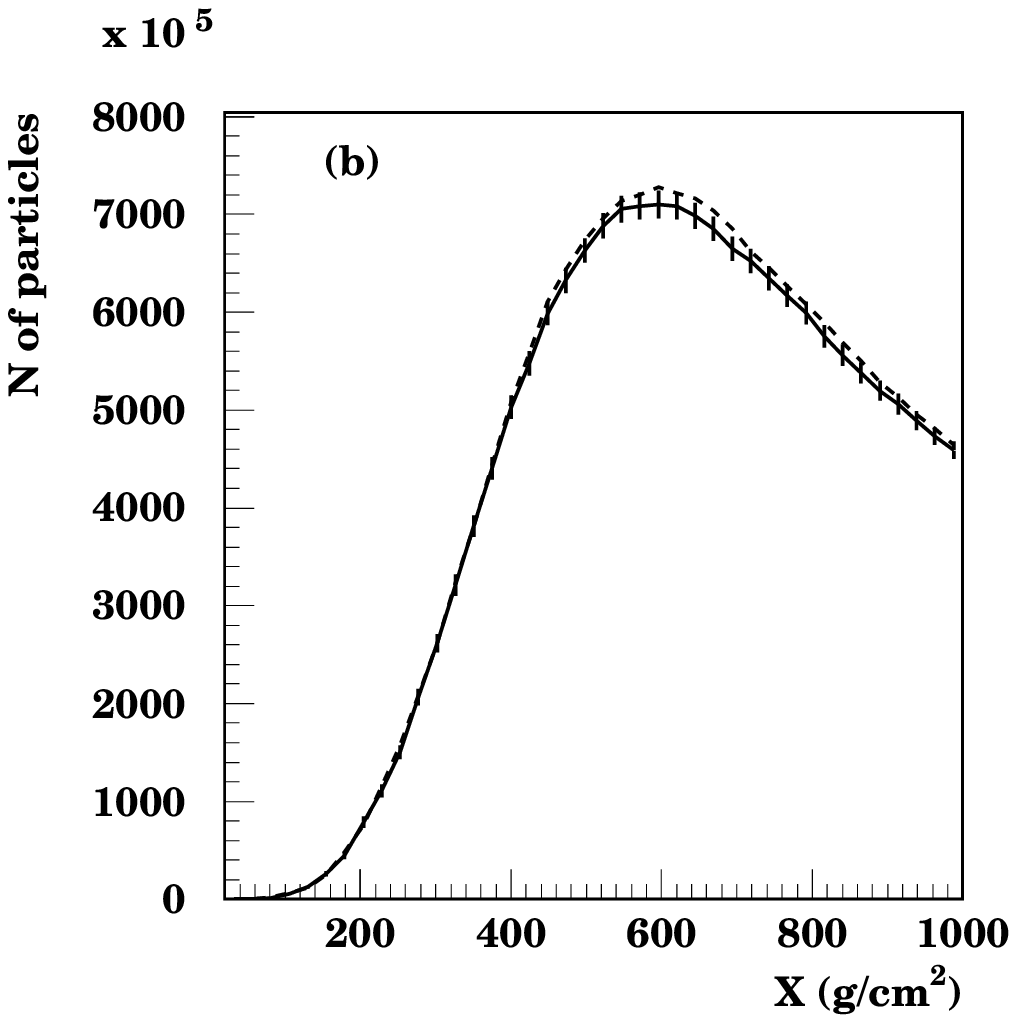}}
\end{array}
\end{displaymath}
\caption{Longitudinal development of charged particles (a) and muons
(b) for showers initiated by proton primaries. The shower parameters
are as in figure \ref{fig:longcharg}.}
\label{fig:longproton}
\end{figure}
\begin{figure}
\begin{displaymath}
\begin{array}{cc}
\hbox{\epsfxsize=7.4cm\epsfig{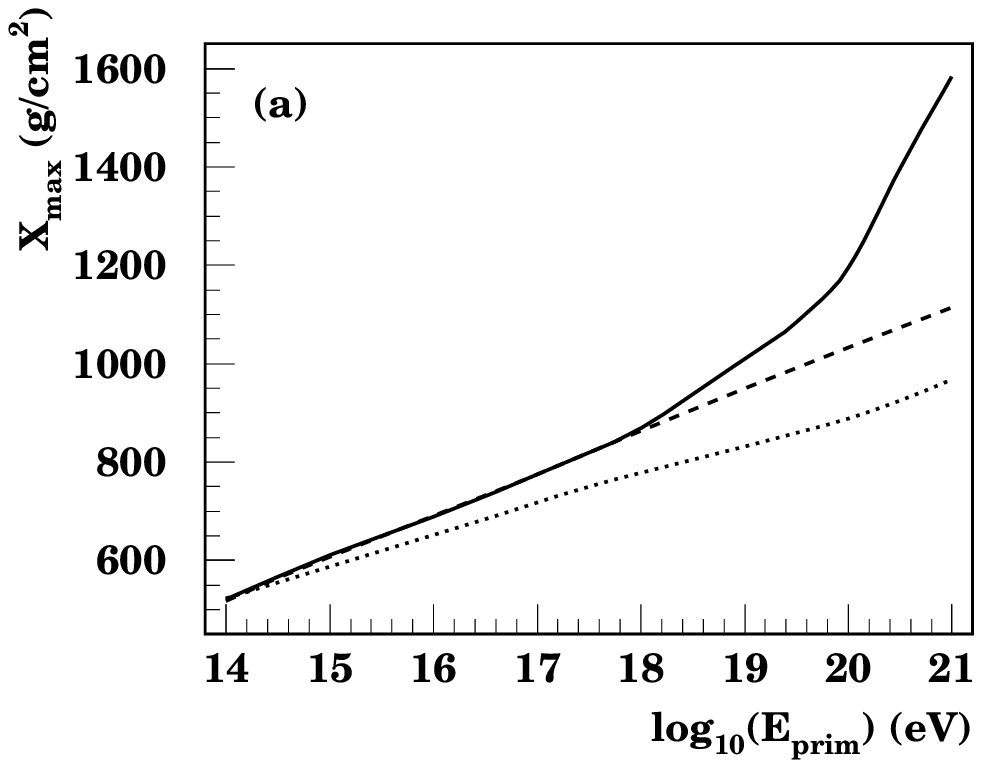}} &
\hbox{\epsfxsize=7.4cm\epsfig{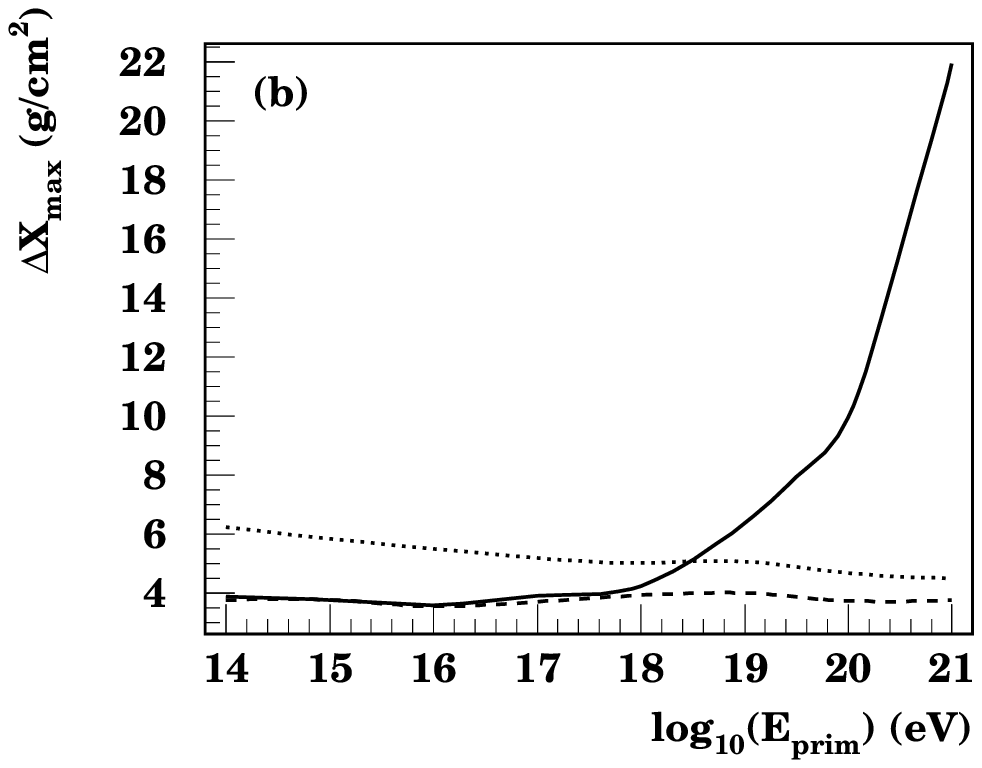}}
\end{array}
\end{displaymath}
\caption{Computer simulation results for the shower maximum, $X_{\rm
max}$, measured along the shower axis (a), and its fluctuations (RMS
error of the mean) (b), plotted as function of the primary energy. The
solid (dashed) lines correspond to gamma primaries taking (not taking)
into account the LPM effect. The dotted lines correspond to proton
primaries.}
\label{fig:xmax}
\end{figure}  
\begin{figure}
\begin{displaymath}
\begin{array}{cc}
\hbox{\epsfxsize=7.4cm\epsfig{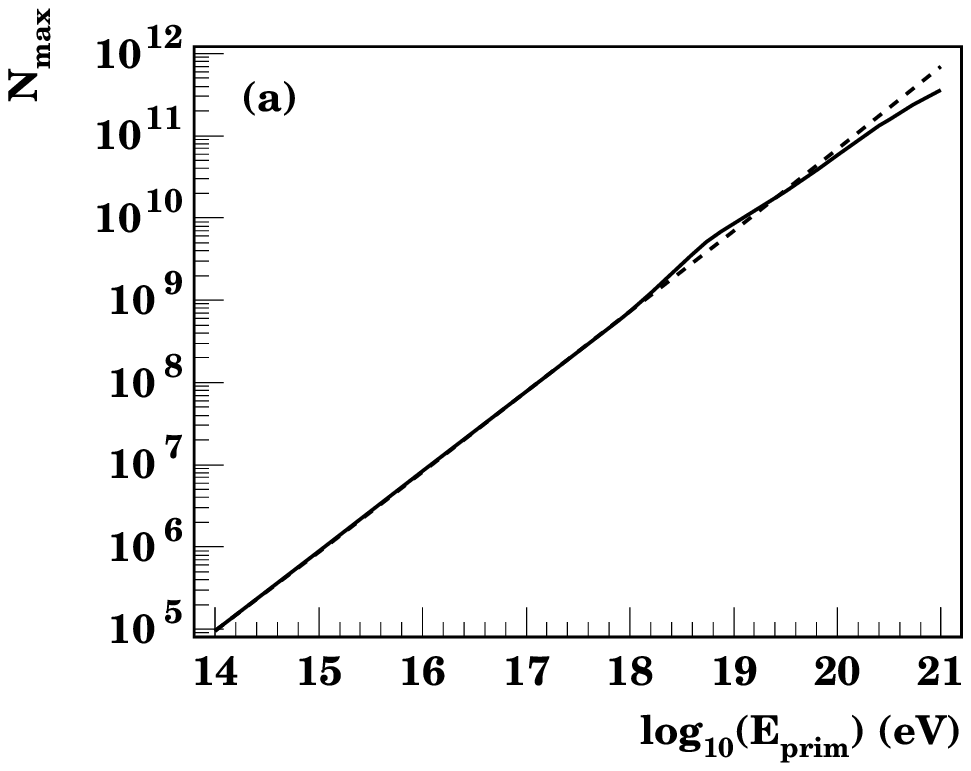}} &
\hbox{\epsfxsize=7.4cm\epsfig{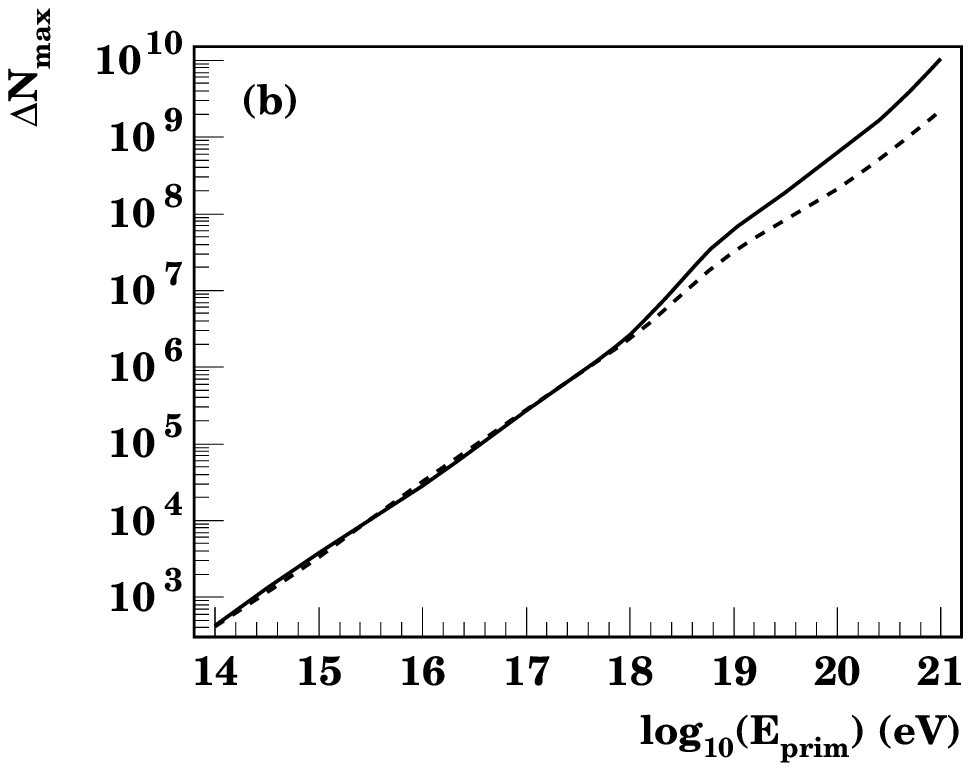}}
\end{array}
\end{displaymath}
\caption{Number of charged particles at the shower maximum, $N_{\rm
max}$, (a) and its fluctuations (RMS error of the mean) (b), plotted
versus the primary energy. The primaries are gammas in the same
conditions as in figure \ref{fig:xmax}.}
\label{fig:nmax}
\end{figure}

\begin{table}
\begin{center}
\begin{tabular}{ccccc}
$\vphantom{A^{B^C}_{B_C}}i$ & $a_i$ & $b_i$ & $c_i$ & $d_i$ \\ \hline
1 &    --   & 7.4783 & --     & 5.0616$\vphantom{A^{B^C}}$ \\
2 &    --   & 30.845 & 11.158 & 11.428 \\
3 & 48.642  & 50.609 & 18.557 & --     \\
4 & 110.12  & --     &  --    & --$\vphantom{A_{B_C}}$ 
\end{tabular}
\end{center}
\caption{The coefficients for the rational expansions of functions $G$ and
$\Phi$ (equations (\ref{G2}) and (\ref{Phi2})).}
\end{table}

\end{document}